\documentclass[conference]{IEEEtran}
\IEEEoverridecommandlockouts
\usepackage{cite}
\usepackage{amsmath,amssymb,amsfonts}
\usepackage{algorithmic}
\usepackage{graphicx}
\usepackage{textcomp}
\usepackage{booktabs}
\usepackage{subfig}
\usepackage{upgreek}
\usepackage{makecell}
\usepackage{nicefrac}
\usepackage[pdftex,dvipsnames]{xcolor}
\def\BibTeX{{\rm B\kern-.05em{\sc i\kern-.025em b}\kern-.08em
    T\kern-.1667em\lower.7ex\hbox{E}\kern-.125emX}}
    
\usepackage[ruled,resetcount]{algorithm2e}
\usepackage{algorithmic}    
\usepackage{lipsum}
\usepackage{xargs}

\newcolumntype{C}[1]{>{\centering\let\newline\\\arraybackslash\hspace{0pt}}m{#1}}
\newcolumntype{R}[1]{>{\raggedleft\let\newline\\\arraybackslash\hspace{0pt}}m{#1}}
\DeclareMathOperator*{\argmin}{arg\,min}

\newcount\colveccount
\newcommand*\colvec[1]{
        \global\colveccount#1
        \begin{pmatrix}
        \colvecnext
}
\newcommand{\colvecnext}[1]{
		#1
        \global\advance\colveccount-1
        \ifnum\colveccount>0
                \\
                \expandafter\colvecnext
        \else
                \end{pmatrix}
        \fi
}
\SetKwRepeat{Do}{do}{while}%
\SetKwInput{KwInput}{Input}
\SetKwInput{KwOutput}{Output}
\SetNlSty{color{gray}}{}{}    
    
\begin{document}

\title{Selfish Energy Sharing in Prosumer Communities:\\A Demand-Side Management Concept\thanks{The author wants to thank the Doctoral Training Alliance (DTA) Energy and Dr. Jean-Christophe Nebel for their support. }%
}

\author{\IEEEauthorblockN{1\textsuperscript{st} Matthias Pilz}
\IEEEauthorblockA{%
\textit{School of Computer Science \& Mathematics} \\
\textit{Kingston University, London}\\
Kingston upon Thames, UK \\
Matthias.Pilz@kingston.ac.uk}
\and
\IEEEauthorblockN{2\textsuperscript{nd} Luluwah Al-Fagih}
\IEEEauthorblockA{\textit{College of Science and Engineering} \\
\textit{Hamad Bin Khalifa University}\\
Doha, Qatar \\
lalfagih@hbku.edu.qa}
}

\maketitle
	
\begin{abstract}
	Global warming is endangering the earth's ecosystem. It is imperative for us to limit green house gas emissions in order to combat rising global average temperatures. One way to move forward is the integration of renewable energy resources on all levels of the power system, i.e.~from large-scale energy producers to individual households. The future smart grid provides the technology for this. \\
	In this paper, a novel demand-side management concept is proposed. It is implemented by a utility company which focuses on renewable energy. Through a specific billing mechanism, prosumers are encouraged to balance load and supply. A game-theoretic approach models households as self-determined rational energy users, that want to reduce their individual electricity costs. To achieve this, they \textit{selfishly} share energy with their neighbours and also schedule their energy storage systems. The scheme is designed such that monetary transactions between households are not necessary. Thus, it provides an alternative approach to energy trading schemes from the literature.
\end{abstract}

\begin{IEEEkeywords}
	Energy Sharing, Game Theory, Smart Grid, Battery Scheduling, Prosumer
\end{IEEEkeywords}

\section{Introduction}
	Climate change is threatening planet earth, its species and people's livelihoods. In an attempt to slow down global warming, greenhouse gas emissions have to be reduced drastically. For instance, instead of burning fossil fuels, energy demands should be fulfiled by renewable energy (RE) resources. This is a challenging task as the intermittent nature of these resources might cause grid instabilities.
	
	 One opportunity to integrate RE into the system is provided by the implementation of smart grids. A smart grid can be described as the combination of a legacy power grid with an additional information technology layer. It is this two-way communication which allows to execute sophisticated energy management systems that can guarantee the stability of the grid.
	
	Furthermore, the smart grid introduces the possibility of a two-way energy exchange. Recently, energy trading has become a buzzword in both industry and academia~\cite{Lee2014,Myung2018,Lee2015,Park2016,Park2017,Pilz2017a,Mediwaththe2018,Zepter2019,Liu2018,Ghosh2018}. The idea to sell excess energy from a private RE resource to a neighbour which is in need, makes the investment in these resources more viable. 
	
	Lee \textit{et al}.~\cite{Lee2014} investigate the relation between small scale (local) producers and end-users. They use a game-theoretic approach to derive a pricing model for direct energy transactions. Their paper belongs to the category of studies (e.g.~\cite{Myung2018,Lee2015,Park2016,Park2017}) which are concerned with a single trading event, i.e.~without a notion of time. Thus, energy storage does not play a role in their analysis.

	An auction mechanism is used in~\cite{Myung2018} to determine the buyer (and the price) of a certain amount of electricity from an individual seller. Their work mainly focuses on the utilisation of smart contracts to automate the process.	
	
	In contrast to~\cite{Myung2018,Lee2015,Park2017}, Park \textit{et al}.~\cite{Park2016} do not use a game-theoretic concept for price determination. Instead, households are `sorted' when they require electricity from their peers, based on historical contributions to the community and the amount they requested. Unfortunately, they do not give insight into how contribution values are influenced by household's decisions. Thus, their idea remains a theoretical concept.
	
	Research combining storage technology and trading of renewable energies began to take shape in 2018 (e.g.~\cite{Zepter2019,Ghosh2018,Mediwaththe2018,Liu2018}). The interest shifted from a one-off trade between households (or microgrids~\cite{Lee2015}), to a planning-oriented approach. Given forecasts for demand and generation (usually for the upcoming day), battery usage and trading activities are scheduled. 
	
	For instance, references~\cite{Zepter2019} and~\cite{Ghosh2018} describe a scenario in which a central operator/platform determines the optimal battery and trading decisions of a community such that the consumption from the external grid is minimised. This does not take into account the preferences of the utility company (UC). While the batteries in these examples are owned by individual households,~\cite{Mediwaththe2018} and~\cite{Liu2018} each employ a single centralised energy storage.
	
	We expect that the most efficient system will have to combine energy exchange between households as well as the utilisation of energy storage. A fully decentralised smart grid might be advertised as the pinnacle of power systems, but it is not the most probable future scenario. We have to acknowledge the role of the UC within the system as they are already investing in large scale RE resources, e.g.~\cite{EWE,EDF,Shell}. 
	
	Furthermore, the energy trading scenarios proposed in the literature assume the willingness of people to initiate monetary transactions every time they exchange energy with one another. We believe that this is overwhelming for the customer. A modern billing scheme needs to be simple and still remain familiar to the existing system to find large scale adoption.
	
	This paper overcomes these shortcomings by proposing a novel demand-side management concept, which has the following advantages:
	\begin{enumerate}
		\item The scheme directly features the generation of energy from the UC and incentivises the community to follow this production.
		\item Each household is treated as an individual and rational entity that wants to minimise their electricity bill. They can do so by scheduling their energy storage system as well as sharing energy with the community.
		\item There are no monetary transactions connected to sharing energy. Nevertheless, it is beneficial for users to offer their excess RE generation.
		%\item The scheme includes energy storage scheduling as well as exchange of energy between individuals in a community.
	\end{enumerate}
	
	The paper is structured as follows: In Section~\ref{sec:system}, the models for the community and its constituents, i.e.~households with battery and RE resource, are introduced. The decisions that can be taken by each household are explained in Section~\ref{sec:decisions}. Section~\ref{sec:utilityFunction} gives details about the billing function, while Section~\ref{sec:game} defines the game in which households aim to minimise their electricity bill. The paper is concluded in Section~\ref{sec:conclusions}.

\section{The System Model}
\label{sec:system}
	In this section, the model for the community is presented. This includes the overall representation of the households and their connections, the demand-side management (DSM) scheme organised by the utility company (UC), the model for the batteries and renewable energy (RE) resources, as well as a detailed description of the decisions each participant can choose from.
	
	\subsection{The Community and DSM scheme}
	\label{subsec:DSMscheme}
	The households in the community are represented as a set $\mathcal{M}$. There are $M=|\mathcal{M}|$ individual houses. Each household $m\in\mathcal{M}$ is a participant of a DSM scheme which is implemented by the UC serving the community. Furthermore, they are all equipped with an RE resource, e.g.~a solar photovoltaic system or a wind turbine, and a lithium-ion battery to store electric energy. Since these households are both producing and consuming energy they are often referred to as \textit{prosumers} in the literature.
	
	The DSM scheme is a process which is repeated once per day. It relies on the two-way communication and power exchange-infrastructure provided by the underlying smart grid. Note that the DSM protocol is automatically executed by the smart meters that are installed in each household. Smart meters are able to measure consumption and generation data of the respective households in discrete intervals $t$. In addition, they also initiate the charging/discharging of batteries as well as the exchange of electricity between the households. The set of all time intervals of the upcoming day is denoted by $\mathcal{T}$ and has $T=|\mathcal{T}|$ elements. Thus the length of a time interval is given by $\Delta t = \nicefrac{24}{T}\,h$.	
	
\begin{figure}
		\centering
		\includegraphics[width=\columnwidth]{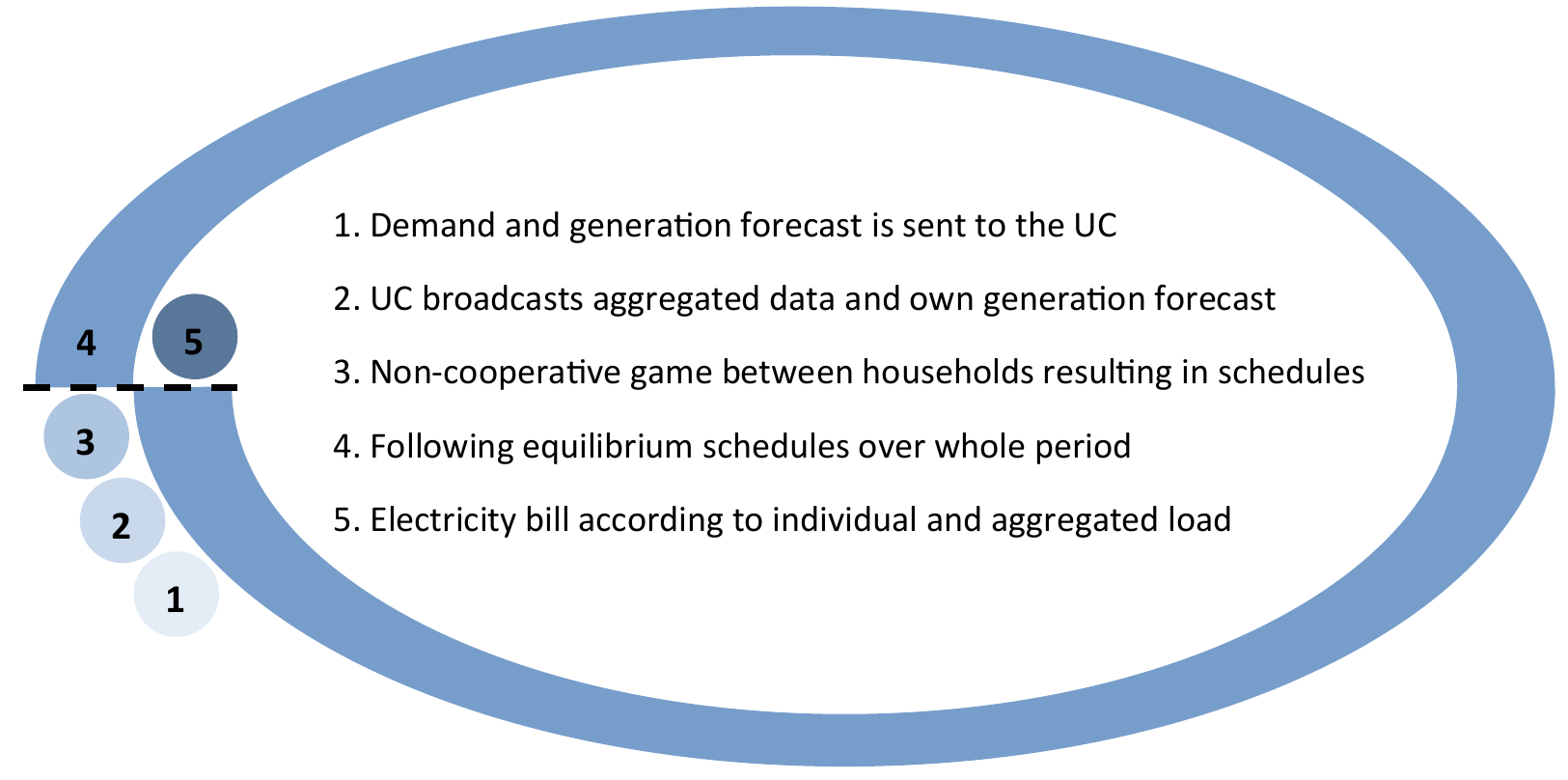}
		\caption{\textit{Demand-side management (DSM) scheme}. The steps that constitute the DSM process are shown schematically. The dashed line indicates the beginning and ending of the scheduling/sharing period.}
		\label{fig:flowchart}
	\end{figure}

	The DSM process performs the following steps (cf.~Fig.~\ref{fig:flowchart}):
	\begin{itemize}
		\item The smart meter of each household executes a forecasting algorithm\footnote{The forecasting is performed based on historical data collected by the smart meter and weather forecasts. The specific algorithm is out of the scope of this paper. In the following it is assumed that this forecast is provided reliably.} to estimate the demand and generation of their respective RE resource. This data is sent to the UC via the communication layer of the smart grid.
		\item The UC aggregates the demand and generation data and broadcasts them to each household. Furthermore, they broadcast their expected electricity generation. Since a considerable amount of their generation also stems from RE resources, these are as well forecasted data.
		\item Based on these information, the households play a non-cooperative dynamic game with the aim to reduce their individual electricity bill\footnote{The billing mechanism implemented by the UC is known to all households in advance.}. The outcome of the game, i.e.~the equilibrium solution, is a collection of instructions. For each interval of the upcoming day it details how to make use of the energy storage system and how energy is exchanged between the households.
		\item During the day, the smart meter of each household follows the equilibrium strategy. Any unilateral deviation from the equilibrium would result in increased costs for the particular user.
		\item The households are billed for each interval according to their individual load as well as the aggregated load of all households. Details on the explicit pricing function are presented in Section~\ref{sec:utilityFunction}.
	\end{itemize}
	
	\subsection{The Household Model}
	\label{subsec:household}
	Households that take part in the DSM scheme own their individual RE resource and a energy storage system.
	
	\subsubsection{Lithium-Ion Battery Model}	
\begin{figure}
		\centering
	\subfloat[charging]{\includegraphics[width=0.48\columnwidth]{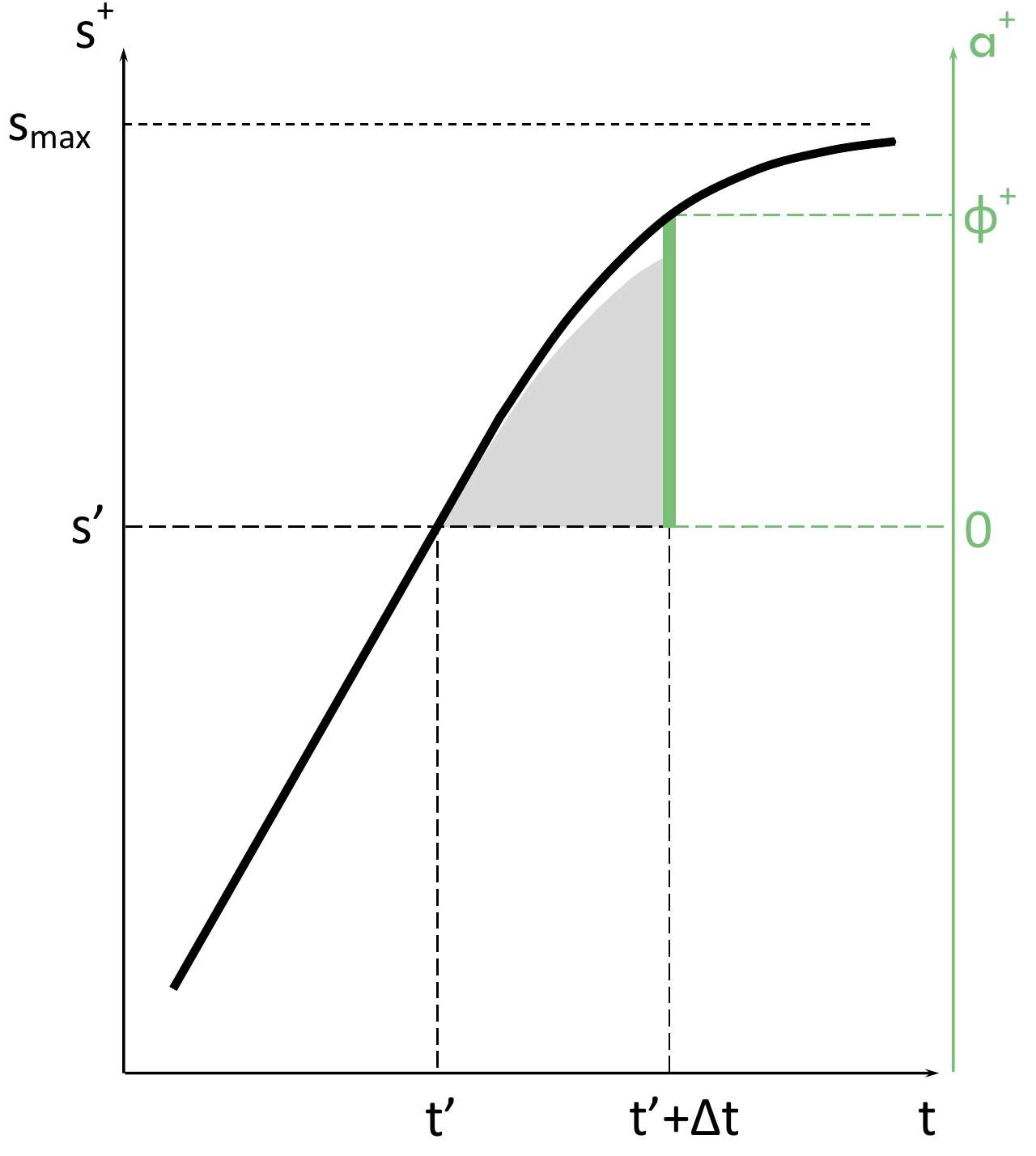}%
	\label{fig:charging}}
	\hfil
	\subfloat[discharging]{\includegraphics[width=0.48\columnwidth]{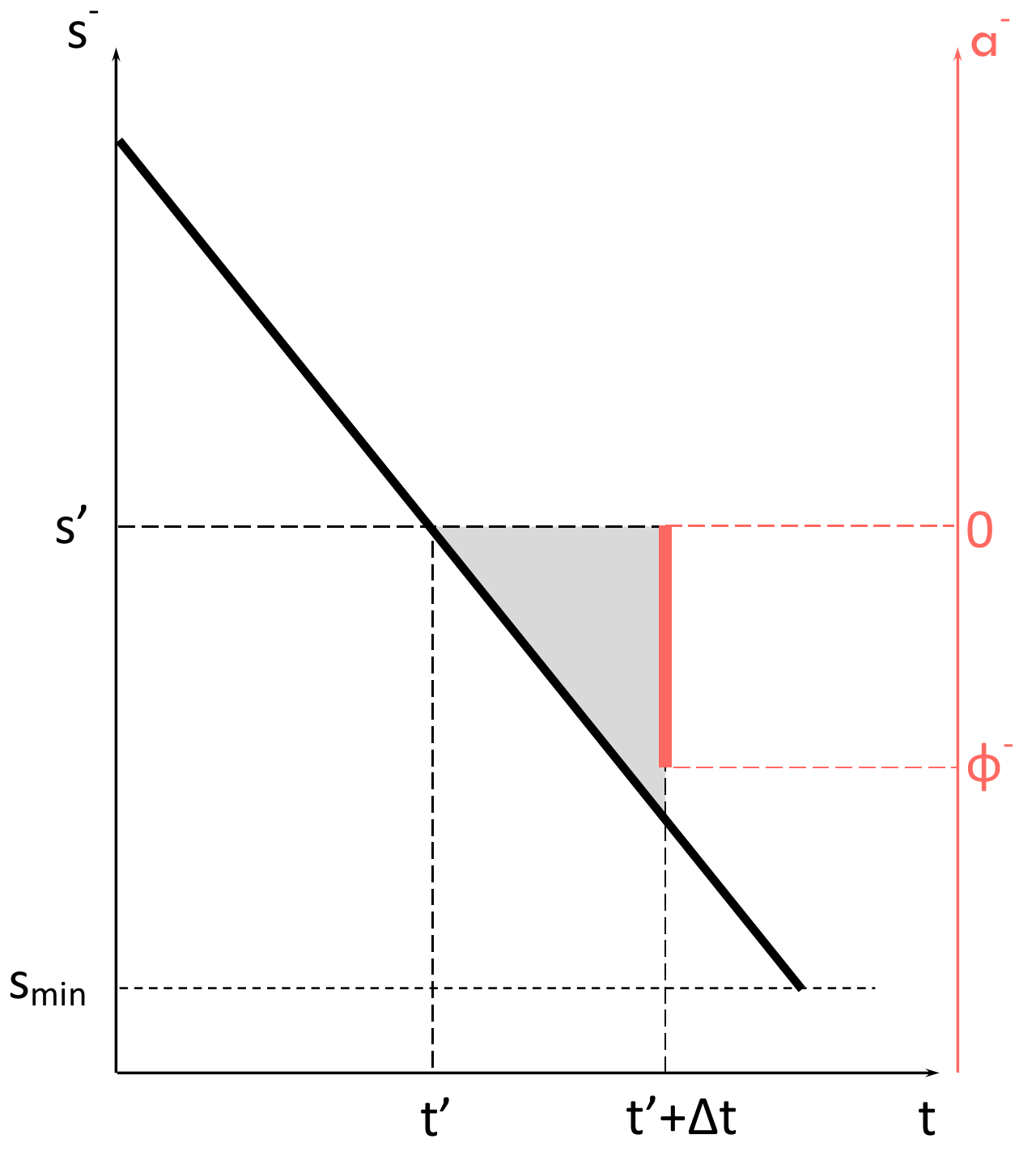}%
	\label{fig:discharging}}
	\caption{\textit{Characteristic charging and discharging behaviour of lithium-ion batteries.} The distinctive charging (discharging) curve is shown on the left (right). At time $t'$ the state-of-charge is given as $s'$. The grey area illustrates the achievable state-of-charge within the following time interval of length $\Delta t$. The left-hand axis denotes the state-of-charge, whereas the right-hand axis represents the charging (discharging) decision with the possible decision interval. The discrepancy between the achievable state-of-charge and the decision interval are due to imperfect efficiencies while charging (discharging). Similar to~\cite{Pilz2018}.}
	\label{fig:batteryModel}
	\end{figure}

	We assume that all storage systems are lithium-ion batteries and implement a detailed model for this technology. The model has been introduced in~\cite{Pilz2017} and was also used in~\cite{Pilz2018a,Pilz2018}. Most notably, the specific charging and discharging characteristics of lithium-ion batteries are considered as shown in Fig.~\ref{fig:batteryModel}.
	
	Charging these type of batteries is done in a two-stage process~\cite{Richtek2014}. The first stage is called the `constant current' (CC) stage. In this stage the state-of-charge (SOC) increases linearly with the charging rate limited to $\rho^+>0$. During the second stage, i.e.~the `constant voltage' (CV) stage, the effective charging rate decreases and eventually levels off towards the maximum capacity $s_\text{max}$ of the battery. The upper bound $\upphi^+(s')$ of how much can be charged within a given time interval of length $\Delta t$ (given an initial value $s'$ for the SOC) is described by
	\begin{displaymath}
		\upphi^+\left(s'\right) = \begin{cases}
					\rho^+\Delta t & \text{if CC charged} \\
					s_{\max}\gamma_1\exp\left[-\frac{\Delta t}{\gamma_2}\right] & \text{if CV charged}		
		\end{cases}\ ,
	\end{displaymath}
	with $\gamma_1$, $\gamma_2$ defined such that the charging curve at the transition point point between the two stages is smooth.
		
	Discharging a lithium-ion battery can be modelled as a linear decrease of the SOC. The slope of this process is defined by the discharging rate $\rho^-<0$. The sharp drop off of the discharging rate, which is usually observed at low capacities~\cite{Richtek2014}, is accounted for by prohibiting discharging below a minimum SOC $s_\text{min}$. We express the limit $\upphi^-(s')$ of how much discharged energy can be used by 
	\begin{displaymath}
		\upphi^-\left(s'\right) = \rho^-\,\Delta t\,\eta_{\text{inv}}\,\eta^-\ ,
	\end{displaymath}		
	which considers the efficiency of the discharging process $\eta^-$ and the efficiency of the DC/AC power inverter $\eta_{\text{inv}}$. The dependency on the current state of charge $s'$ is implicitly considered due to the minimum SOC $s_\text{min}$.
	
	The battery model also includes self-discharging, whenever the battery remains idle for an interval $\Delta t$. This process is characterised by the self-discharging rate $\bar{\rho}<0$.
		
	\subsubsection{Renewable Energy Resource Model}	
	For simplicity, we only consider local installations of wind turbines and solar photovoltaic (PV) panels; one of the two for every household. The smart meters have the capability to forecast the output of the respective energy resource for the upcoming day (cf.~\cite{Rana2016,Bichpuriya2016,Dolara2015} for suitable forecasting algorithms). For a time interval $t$ this output is denoted by $w^t$. Within our model, there are three possibilities how this energy can be used: 
	\begin{itemize}
		\item Direct usage by household appliances 
		\item Charging the battery of the respective household 
		\item Sharing the energy with other households 
	\end{itemize}

	When the energy is used directly by household appliances it needs to be converted from DC to AC with an efficiency of $\eta_{\text{inv}}$. This conversion is not necessary when the energy is used to charge the battery. In the case of sharing the energy with other users, we consider DC/AC conversion as well as line losses. 
	
	Note that direct usage is always prioritised. This means that charging one's own battery or sharing energy can only be performed if the produced amount exceeds the demand during the respective time interval. More explicitly, we consider the net-demand $d^t_m$ of household $m$ at time interval $t$ as an indicator for the available usage options:
	\begin{equation}
		d^t_m = \bar{d}^t_m - \eta_{\text{inv}}\,w^t_m\ ,
		\label{eqn:netDemand}
	\end{equation}
	where $\bar{d}^t_m$ is the actual demand caused by the household appliances.
	
\section{Decision Variables}
\label{sec:decisions}
	In this section, we introduce the decisions that can be taken by the households. It will become clear how the battery and renewable energy (RE) resource model (cf.~Section~\ref{sec:system}) limit these choices. 
	
	In general, there are two types of decisions to make for each household $m\in\mathcal{M}$ during each interval $t\in\mathcal{T}$ of the upcoming day: 
		\begin{itemize}
			\item how to use the battery (denoted by $a_m^t$)
			\item how to share energy (denoted by $e_m^t$)
		\end{itemize}			
		For simplicity we combine these into a single decision vector $x_m^t=(a_m^t,\,e_m^t)$. A collection of these decisions over the entire time period $x_m = (x_m^1,\,x_m^2,\,\dots,\,x_m^T)$ is called a schedule. 
		
		For each time interval, households are categorised as \textit{giver} or \textit{taker} of energy based on their net-demand~\eqref{eqn:netDemand}. The classification in giver and taker introduces different boundaries to the decisions $x_m^t$. 
		
	\subsection{Definitions of Giver and Taker}
	\label{subsec:defGiverTaker}
	If $d^t_m>0$ it means that the demand of the household cannot be satisfied from their own RE production. Thus they require further electricity from either the grid or other households. We call such a household \textit{taker}.
	
	If $d^t_m\leq 0$ it means that there is an excess of RE which can be used to charge their own battery or be shared with the other households. During such an interval the household is categorised as a \textit{giver}. 
	
	The load $l^t_m$ of a user is defined as the amount of energy drawn from the grid, i.e.~from the utility company. In the following section (Section~\ref{sec:utilityFunction}), the explicit billing mechanism highlights how the electricity costs depend on the aggregated load of all households as well as their individual load. Here, we explain how the decisions taken by a household affect the load.

	\begin{figure}
		\centering
		\includegraphics[width=\columnwidth]{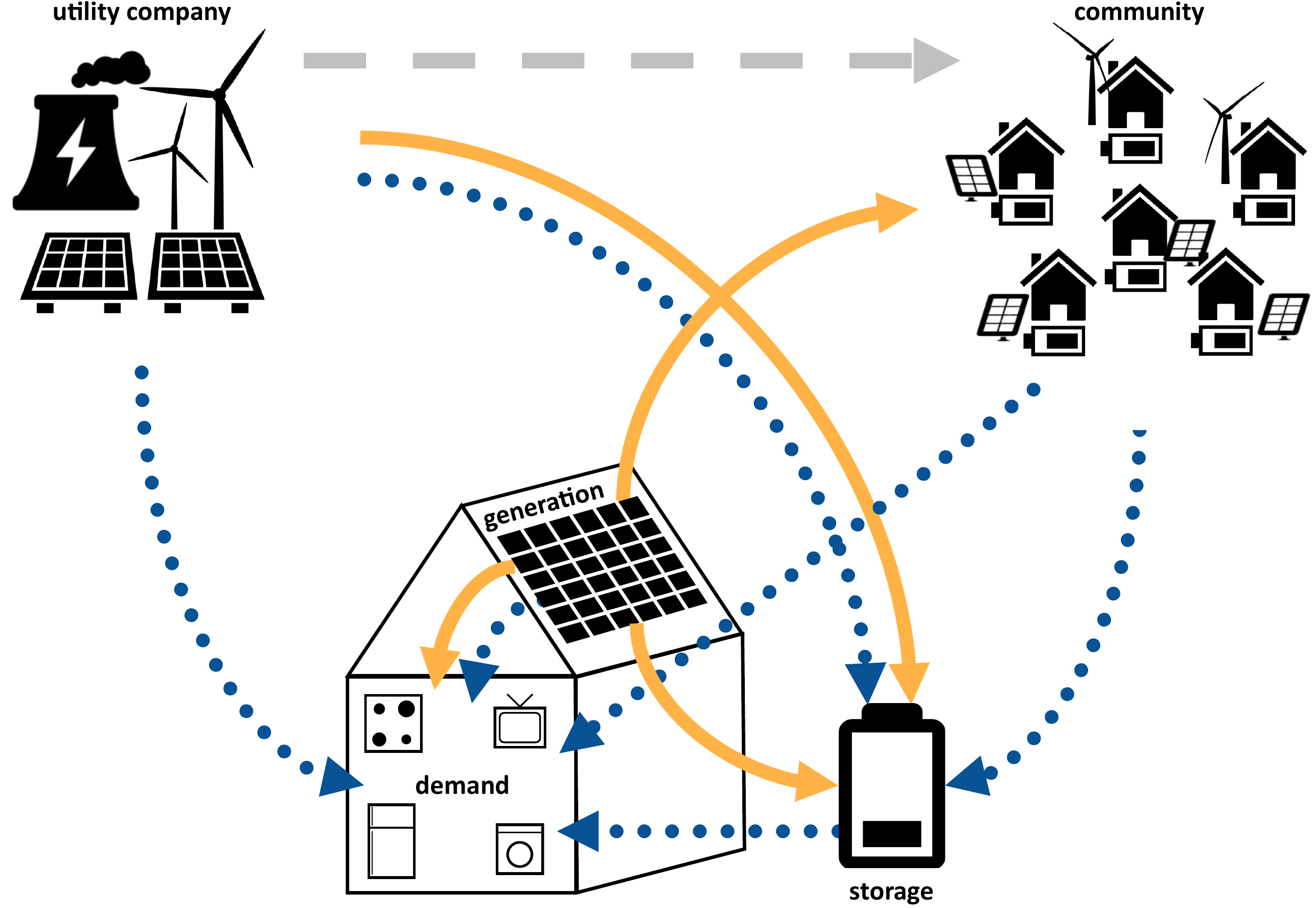}
		\label{fig:decisionArrows}
		\caption{\textit{Electricity flows}. The potential electricity flows for an individual household of the prosumer community is shown. Dotted lines refer to flows of a `taking' household, while solid lines reference a `giving' household. These flows are also valid for all the other households in the community and summarised in the dashed line between the utility company and the community.}
	\end{figure}	
	
	\subsection{Decision Space for `Taking' Households}
	\label{subsec:decisionTaker}
		The load of a taking household $m\in\mathcal{M}$ at time $t\in\mathcal{T}$ is 
		\begin{equation}
			l_m^t = d^t_m + a^t_m + e^t_m\ .
			\label{eqn:loadTaker}
		\end{equation}
		The battery decision $a^t_m$ and sharing decision $e^t_m$ have to fulfil the following constraints:
		\begin{subequations}
			\begin{equation}
				s^t_m-s_{\text{min}} < \upphi^-\left(s^t_m\right) \leq a^t_m \leq \upphi^+\left(s^t_m\right) < s_{\text{max}}-s^t_m
				\label{subeqn:batTaker}
			\end{equation}
			\begin{equation}
				-d^t_m-a^t_m \leq e^t_m \leq 0
				\label{subeqn:shareTaker}
			\end{equation}
			\begin{equation}
				\left| e^t_m \right| \leq E^t\ .
				\label{subeqn:poolTaker}
			\end{equation}
		\end{subequations}
		Equation~\eqref{subeqn:batTaker} describes the discharging and charging restrictions from left to right, respectively. Discharging cannot be done below the minimum SOC $s_\text{min}$ and also the charging rate and efficiency are respected. The right hand side makes a similar statement for the charging process (cf.~Section~\ref{subsec:household}). Note that discharging refers to using energy from the battery to fulfil demand, while charging the battery explicitly refers to charging the battery from the grid.
		
		Equation~\eqref{subeqn:shareTaker} guarantees $l^t_m \geq 0$, meaning that the households can neither sell energy from their battery nor the energy they receive from their neighbours to the grid. Furthermore, the amount of energy they can obtain from the neighbours is restricted to the excess production of them at that time interval (cf.~\eqref{subeqn:poolTaker}). Note that charging the battery can effectively be done from shared energy.
		
		The interaction with the battery directly influences the SOC. To reflect the correct state $s^{t+1}_m$ of the upcoming interval we calculate
		\begin{displaymath}
			s^{t+1}_m = \begin{cases}
				s_m^t + \eta_{\text{inv}}\;\eta^+ a^t_m &\ , \text{if } a^t_m>0 \\
				s_m^t + \nicefrac{a_m^t}{(\eta_{\text{inv}}\;\eta^-)} &\ , \text{if } a^t_m<0 \\
				s_m^t\cdot\left(1 + \bar{\rho}\right)^{\Delta t} &\ , \text{if } a^t_m=0
				\end{cases}\ .
		\end{displaymath}
	
	\subsection{Decision Space for `Giving' Households}
	\label{subsec:decisionGiver}
	The load of a giving household $m\in\mathcal{M}$ at time $t\in\mathcal{T}$ is 
		\begin{equation}
			l_m^t = a^t_m\ .
			\label{eqn:loadGiver}
		\end{equation}
		All the demand $\bar{d}_m^t$ is already fulfilled from their RE resource. Nevertheless, there is also a sharing decision to make: The household can offer (part of) their excess production $d_m^t$ to the community, i.e.
		\begin{equation*}
			E^t \leftarrow E^t+\bar{\eta}\cdot e^t_m\ ,\ \ \text{with }\ 0 \leq e^t_m \leq -d^t_m\ ,
			\label{eqn:sharGiver}
		\end{equation*}
		where $0<\bar{\eta}\leq 1$ denotes the line losses of the network. The remaining part, i.e.~$-d^t_m-e^t_m$, is automatically used to charge the battery. This gives a further restriction to the amount to be charged from the grid as follows:
		\begin{equation*}
			0 \leq a^{t}_m-d^t_m-e^t_m \leq \upphi^+(s^t_m) < s^{\text{max}}_m-s^t_m\ .
			\label{eqn:charGiver}
		\end{equation*}
		In the case that the effective charging amount is exactly 0, the self-discharging process of the battery $s^{t+1}_m = s_m^t\cdot\left(1 + \bar{\rho}\right)^{\Delta t}$ is considered (cf.~Section~\ref{subsec:decisionTaker}). Otherwise the SOC for the upcoming interval is calculated to be
		\begin{displaymath}
			s^{t+1}_m = s^t_m + \eta_{\text{inv}}\;\eta^+ a_m^t + \eta^+\cdot(-d^t_m-e^t_m)\ .
		\end{displaymath}
		When charging the battery directly from locally produced RE resources, only the battery efficiency $\eta^+$ has to be considered, whereas charging from the grid also requires conversion from AC to DC with the respective conversion efficiency $\eta_{\text{inv}}$.

\subsection{The Aggregated Load}	
\label{subsec:aggLoad}
	The aggregated load $L^t$ at time interval $t$ is the total amount of electricity requested by the community. With respect to the non-cooperative game (cf.~Section~\ref{sec:game}) the following definition is used:
	\begin{equation}
		L^t = l^t_m + l^t_{-m}\ ,
		\label{eqn:aggLoad}
	\end{equation}
	where $l^t_m$ is the load of a specific household $m\in\mathcal{M}$ and $l^t_{-m}$ is the load of all households except for $m$. 

	\begin{figure*}[t]
		\centering
		\includegraphics[width=0.9\textwidth]{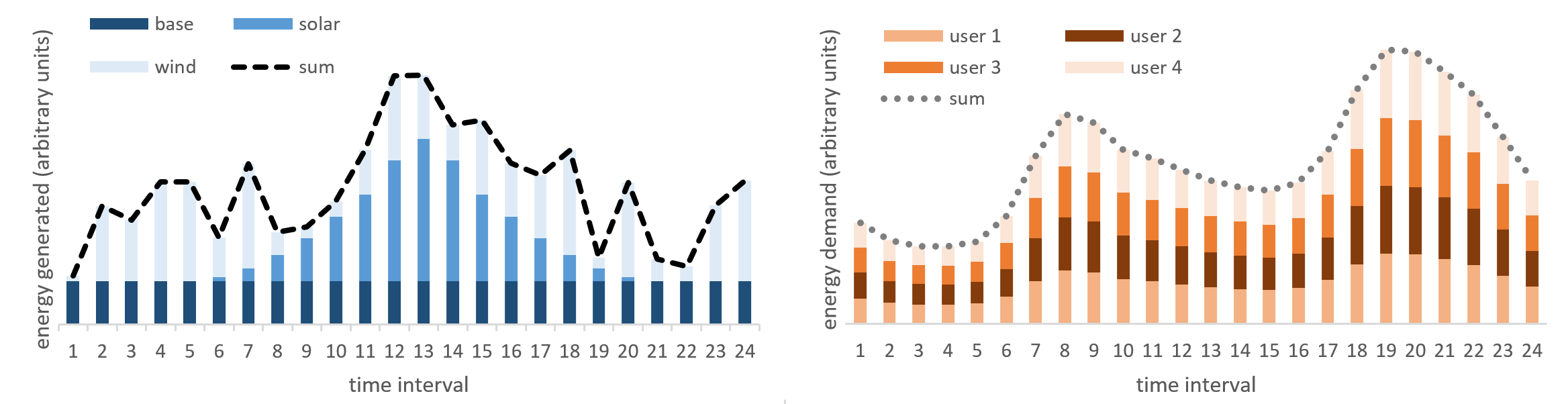}
		\caption{\textit{Generation and demand}. The exemplary energy generation of the utility company and demand of four households is shown on the left and right, respectively. Furthermore, the sum of the individual components are shown as lines. \newline $g^t$ refers to the dashed line in the left-hand plot. The dotted line in the right-hand plot can be seen as the aggregated load $L^t$ in a scenario where batteries are not used and energy is not shared.}
		\label{fig:genDem}
	\end{figure*}		
	
	In the previous subsections it was shown how each household is able to directly influence their own load, i.e. Equations~\eqref{eqn:loadTaker} and~\eqref{eqn:loadGiver}. Furthermore, it became clear that by means of the sharing ability they can also (indirectly) affect the load of other households in the community.

\section{The Utility Function}
\label{sec:utilityFunction}
	In this section, the novel utility function, i.e.~the electricity bill, for each participant of the game is presented. %To do so, the pricing strategy imposed by the UC is introduced first. 

	Households are incentivised to follow the forecasted production pattern $g^t$ of the utility company (cf.~Fig.~\ref{fig:genDem}) by implementing the following price per energy unit for a given interval $t\in\mathcal{T}$ of the upcoming day:
	\begin{equation*}
		p^t = \left(L^t - g^t\right)^2 + p_0\ ,
		\label{eqn:price}
	\end{equation*}
	where $p_0>0$ is constant and $L^t$ is the aggregated load of all users as defined in~\eqref{eqn:aggLoad}. The closer the aggregated load $L^t$ is to the generated electricity $g^t$, the smaller the price per energy unit for this particular interval $t\in\mathcal{T}$. The electricity bill for a particular household $m\in\mathcal{M}$ for one day is then calculated to be:
	\begin{equation}
		U_m = \sum_{t=1}^T l^t_m\cdot p^t\ .
		\label{eqn:bill}
	\end{equation}
	In the non-cooperative game between the households, each user strives to minimise their individual electricity bill~\eqref{eqn:bill}. In the interest of showing how the electricity bill for household $m\in\mathcal{M}$ depends on their own sharing/charging decisions $x_m$ (cf.~Section~\ref{sec:decisions}) and the decisions $x_{-m}$ of all the other households 
	%$-m=\{m'\in\mathcal{M}\,|\,m'\neq m\}$ 
	let us rewrite~\eqref{eqn:bill} explicitly with these dependencies:
	\begin{align}
		U_m\left(x_m,x_{-m}\right)& = \sum_{t=1}^T\biggl\{l^t_m(x_m)\cdot p_0 \label{eqn:billLong}\\
		 &+ l^t_m(x_m)\cdot\left[l^t_m(x_m) + l^t_{-m}(x_{-m}) - g^t \right]^2 \biggr\}\ . \notag
	\end{align}

\section{The Non-Cooperative Game}
\label{sec:game}
	In this section, we define the non-cooperative game and explain the solution approach that leads to equilibrium schedules for the individual households. 
	
	The non-cooperative game is defined by $G=\left\{\mathcal{M},\mathcal{X},U\right\}$ with
	\begin{itemize}
		\item $\mathcal{M}$ as the set of participants of the game.
		\item $\mathcal{X}=\mathcal{X}_1\times\cdots\times\mathcal{X}_M$, where $\mathcal{X}_m$ is the set of all actions $x_m$ that fulfil the restrictions detailed in Section~\ref{sec:decisions}.
		\item $U=\left(U_1,\dots,U_M \right)$, with the utility function~\eqref{eqn:billLong}\newline ${U_m:\mathcal{X}\rightarrow{\rm I\!R}}$ for player $m$.
	\end{itemize}
	An iterative best-response algorithm~\cite{Shoham2009} is used to solve the game. The solution is a vector of battery/sharing-schedules ${\hat{x}=\left(\hat{x}_m,\hat{x}_{-m}\right)}$, i.e.~one for each participant\footnote{Equivalently, we can write the solution as ${\hat{x}=\left(\hat{x}_1,\dots,\hat{x}_M\right)}$ to emphasis the contribution of all households.}. 	
	
During each step of the iteration, the households determine their best strategy by solving the minimisation problem:
	\begin{equation}
		\hat{x}_m = \argmin_{x_m\in \mathcal{X}_m} U_m\left(x_m,x_{-m}\right)\ .
		\label{eqn:Uargmin}
	\end{equation}
A summary of the algorithm can be found in~Algorithm~\ref{alg:bestResponse}. 
\begin{algorithm}[h]
  \caption{Best-response algorithm for finding a pure NE based on \cite{Shoham2009}}
    \label{alg:bestResponse}\KwInput{$g^t$, $w^t$, $\bar{d}^t$} 
\kern-6pt
\hrulefill\\
		initialise random vector of schedules $x=(x_m,x_{-m})$ \\
%		\nextnr\label{alg:BR_while}	
		\While{there exists a player $m$ for whom $x_m$ is not a best response to $x_{-m}$}{
%		\nextnr\label{alg:BR_forN}	
		\For{\normalfont \textbf{each} $m\in\mathcal{M}$}{
			\begin{tabular}{cl}
				$\hat{x}_m$ & $\leftarrow$ best response to $x_{-m}$ based on~\eqref{eqn:Uargmin}\\
        		$x$ & $\leftarrow \left(\hat{x}_m,x_{-m}\right)$	
			\end{tabular}
		}		    
	}
\kern-6pt
\hrulefill\\
\KwOutput{$\hat{x}=x$}
\end{algorithm}
When this iteration procedure terminates, each household has determined a schedule $x_m$ which is in equilibrium with all the other households. This means that there is no incentive to deviate from this strategy. Any unilateral deviation can only ever result in a worse outcome, i.e.~a more expensive electricity bill.

\begin{figure}[t]
		\centering
		\includegraphics[width=0.8\columnwidth]{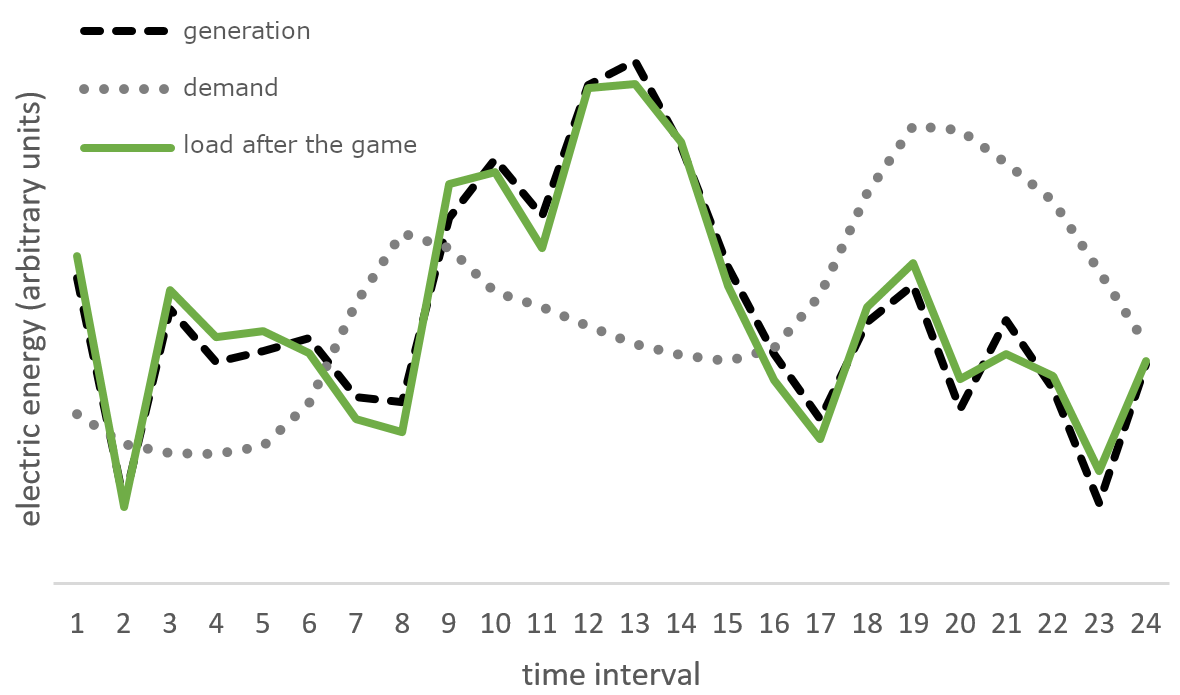}
		\caption{\textit{Outcome of the game}. The dashed and dotted lines are directly taken from Fig.~\ref{fig:genDem} and refer to the generation of the utility and the demand of the community, respectively. The utility function~\eqref{eqn:bill} is designed such that it is most beneficial for the households to make use of the energy storage systems and share energy in a way that the aggregated load matches the forecasted supply. The solid line indicates the potential aggregated load of the households after playing the game.}
		\label{fig:result}
	\end{figure}

Fig.~\ref{fig:result} illustrates the potential effect of the scheme if it were to be applied on the data shown in Fig.~\ref{fig:genDem}. By scheduling their batteries and sharing energy among each other, they are able to adapt to the forecasted generation of the UC.

\section{Conclusions}
\label{sec:conclusions}
The adoption of renewable energy resources helps to limit greenhouse gas emissions. This paper proposed a demand-side management scheme which allows the integration of renewable energy resources on both the utility company's level (\textit{large scale)} and the customer's level (\textit{small scale)}. Within the scheme, households are financially incentivised to adjust their load to the forecasted electricity production of the utility company. They can accomplish this by scheduling their locally installed energy storage systems and by sharing energy with the community. The underlying process to organise the scheduling and sharing is based on a non-cooperative game in which every participant is only interested in achieving the best for themselves. We thus established a mechanism for \textit{selfish energy sharing} which does not require direct monetary transactions between the prosumers. This makes it more approachable for the user, fostering wide-spread adoption.

Future research will quantify the gains of our approach in terms of renewable energy self-consumption, how closely the utility's production curve can be followed, and financial rewards for the prosumers. Furthermore, a detailed comparison to energy trading schemes will be necessary to demonstrate the competitiveness of our proposition.
It is worth noting that since the optimal size of storage and generation for each household is derived using the billing scheme and energy trading model, their optimality has to be revisited for all of these demand-side management concepts.

%%%%\bibliography{./bib_SGC}

\end{document}